\documentclass[a4paper,11pt]{article}
\usepackage{pos}

\title{On probing self interacting dark matter models through the absorption of gravitational waves}
\ShortTitle{On probing SIDM through the absorption of GWs}

\author*[a]{Víctor Fonoll}

\affiliation[a]{Departamento de Física Teórica, Universidad de Zaragoza,\\
  C. Pedro Cerbuna 12, 50009 Zaragoza, Spain}

\emailAdd{vfonoll@unizar.es}

\abstract{In the forthcoming years, the study of the fundamental interactions between gravitational waves (GWs) and matter will be crucial in order to understand what the new generations of GWs detectors will tell us. We present the inverse bremsstrahlung (IB) absorption of GWs as a novel approach to GWs physics that can help set constraints on different physical models. We study the absorption of GWs in scattering processes of interacting dark matter. The observation of GWs of a given frequency sets constraints on its absorption efficiency. In the case of interacting dark matter, this can translate to constraints on its mass-coupling space, or in its temperature. For this, we parametrize the absorption of GWs in DM halos and in IGM, at low and very high redshifts. We find the arising constraints to be less stringent than existing ones.}

\FullConference{3rd General Meeting of the COST Action COSMIC WISPers (CA21106) - 3rd Training School of the COST Action COSMIC WISPers (CA21106) (COSMICWISPers2025)\\
9–12 Sept 2025 and 16–19 Sept 2025\\
Sofia, Bulgaria and Annecy, France\\}

\begin{document}
\maketitle

\section{Introduction}
Gravitational waves can be absorbed via inverse Bremsstrahlung in scattering processes \cite{iBs}. The detection of GWs of a particular frequency implies then that the Universe cannot be opaque to such frequencies - in other words, that the medium in which GWs have travelled cannot absorb efficiently GWs of such frequency. This is parametrized by the optical depth $\tau$, 
 \begin{equation}\label{tau}
     \tau = \int_{D}\text{d}s\;\Gamma_{\text{abs}}(s)/c
 \end{equation}
 which tells us about the opacity along $D$ for a given rate of absorption \textit{per graviton} $\Gamma_{\text{abs}}$. Similarly to the study of $\tau$ in the context of CMB photons and the reionization history of the Universe, a consistent study of the interaction of GWs with its medium of propagation could help us to study it.\\
 
 As a first step towards this goal, we will consider the hypothetical direct detection of GWs in the context of self-interacting dark matter (SIDM) $-$ see e.g. \cite{review} for a review on SIDM. We will see that $\tau$ is parametrized in terms of the DM self-couplings, the mass, the temperatures and the number densities. Then, the detection of GWs sets constraints on those parameters by setting constraints on the optical depth. As a toy model, we will do this in the context of ultra-light bosonic DM (ULDM) with a quartic coupling \cite{BEC, ULDM}.
\section{Absorption of GWs in ULDM self-scattering}
 The first ingredient we need is the rate of absorption of a graviton in a medium. For a non-relativistic 2-2 scattering,
 \begin{equation} \label{AbsortionRate}
    \Gamma_{\text{net abs.}} = \frac{G \mu^2}{5\pi^2\hbar c^2  f^3} n_1 n_2 \overline{v^5 \sigma_D}\Big[\frac{2\pi\hbar  f}{k T}\Big],
\end{equation}
where  $n_1$ and $n_2$ are the number densities of the two colliding particles, $\mu$ is their reduced mass, $v$ their relative velocity, an overline denotes a thermal average, and 
$ \sigma_D = \int \left(\frac{d\sigma}{d\Omega'}\right)\sin^2\theta_c  \;d\Omega'$ \cite{iBs}.
The last expression refers to the differential cross-section \textit{without} gravitons.\\

Then, we will use this expression for the case of our particular model, a scalar field with a quartic self-coupling with masses in the range $10^{-24}$ eV $ < m < 1$ eV.\; For a Lagrangian $ \mathcal{L}_{\text{int}} = -\lambda\;\psi^4/4!,$
the cross section in the NR limit is $\text{d}\sigma/\text{d}\Omega=\lambda^2/256\pi^2m^2,$
where $m$ is the mass of the DM particle.
Then, it is straightforward to find the rate
\begin{equation}\label{UL}
    \Gamma_{\text{abs}} = 1.13\cdot10^{-27} \frac{1}{\text{s}}\frac{\lambda^2[n_(\text{cm}^{-3})]^2}{[ f(\text{Hz})]^3}\left(\frac{kT}{mc^2}\right)^{5/2}\Bigg[\frac{2\pi\hbar f}{kT}\Bigg].
\end{equation}
\section{Characterization of the mediums of propagation}
The rate at which gravitons are absorbed depends on the number density and temperature of DM. Then, in order to compute the optical depth $\tau$, we need to characterize the medium in which the absorption takes place. 
\subsection{Absorption in DM halos}
We consider first a graviton that comes across a single halo of mass $M_h$ virialized at $z_{vir}$. For a spherical halo, the diameter will be $
    D_{h}(z_{vir}, M) =3\left(M/\rho_h(z_{vir})\right)^{1/3}/2\pi,$
where we estimate the density of the halo as $\rho_h (z_{vir }) = 200\rho_c(z_{vir})$, being $\rho_c$ the critical density. Taking $H(z) \approx H_0\; \Omega_{m,0}^{1/2}(1+z)^{3/2}$, $H_0 = 67\; \text{km/s/Mpc}$, $\Omega_0 = 0.3$, one can parameterize the virial temperature of a particle of mass $m$ as
$
     T_{vir} (z_{vir}, M) = 5\cdot 10^{-3} \;[m \;(\text{GeV})] \left[{M_h (M_\odot)}\right]^{2/3}(1+z_{vir}) \;\text{K}.$  \cite{MBW}
From expression \eqref{tau},  assuming that the distance traveled is $D_h$ and the conditions are homogeneous, one finds the optical depth of a single halo to be $\tau_h(z_{vir}, M) = \Gamma(z_{vir}, M)\cdot D_{h}(z_{vir}, M)/c,$ where we emphasize that everything is parametrized in terms of $z_{vir}$ and $M$.\\
\vspace{-0.2 cm}

Our goal is to compute the optical depth due to DM halos for a graviton emitted at $z_i \sim 30$, when structure formation becomes relevant. For this, $\tau_h$ gives the contribution to the optical depth of a halo of mass $M_h$ encountered at $z \approx z_{vir}$. We can mildly expect the total optical depth to be  $\tau = \int \tau_h dN_h$ $-$ this is, the sum of optical depths of all halos encountered. To find $\text{d}N_h$, the differential number of halos that the GW will come across since $z_i$ until today is 
\begin{equation} 
    \begin{aligned} \label{dNdM}
        \frac{dN_h(z,M)}{dM} = \int \frac{dn_h(z,M)}{dM}\text{d}V(z,M)
        = \int \frac{dn_h(z,M)}{dM}\frac{c\;S_h(z,M)}{H(z)(1+z)} \text{d}z ,
    \end{aligned}
\end{equation}
where  $\text{d}n_h/\text{d}M$ is the halo mass function HMF \cite{PS}, the comoving differential number density of galaxies at a given redshift in the mass range $\left(M, M + \text{d}M\right)$. We have also taken $\text{d}V = S_h(z,M) |c\text{d}t/dz| \text{d}z$, where $S_h$ is the surface of the halo projected in the line of sight $S_h = \pi R_h^2$, with $R_h = D_h/2$. Then, from $\tau_h$ and \eqref{dNdM}, the total optical depth for a graviton emitted at $z_i$ encountering halos of masses $M_{\text{min}} < M < M_{\text{max}}$ is
\begin{equation}\label{TauPS}
     \tau_{\text{halos}} = \int_{0}^{z_i} \int_{M_{\text{min}}}^{M_{\text{max}}}\frac{\Gamma_{\text{abs}}(z,M)D_h(z,M)S_h(z,M)}{H(z)(1+z)}\frac{dn_h}{dM}(z,M) (1+z)^3\; \text{d}M\text{d}z.
\end{equation}
\subsection{Background absorption}
We will also consider the absorption of GWs by non-virialized matter in IGM. For DM particles, we consider the background quantities (for $z\gg1$)  $\rho_{DM}(z) = \Omega_{DM}(z)\rho_c(z) \approx \Omega_{DM,0}\;\rho_{c,0}(1+z)^3$ and $T_{DM}(z) = T_{DM,0}(1+z)^2$. With this, $\Gamma_{abs}$ is parametrized simply in terms of the redshift and
\begin{equation}
     \tau_\text{{bg}} = \int_0^{z_{em}} \frac{\;\Gamma_{\text{abs}}(z)}{H(z)} \text{d}z .
\end{equation}
\vspace{-0.8 cm}

\section{Prospects for constraints on self-interacting DM}
\vspace{-0.2 cm}

\begin{figure}[h!]
\centering
\begin{minipage}{.5\textwidth}
  \centering
  \includegraphics[width =\linewidth]{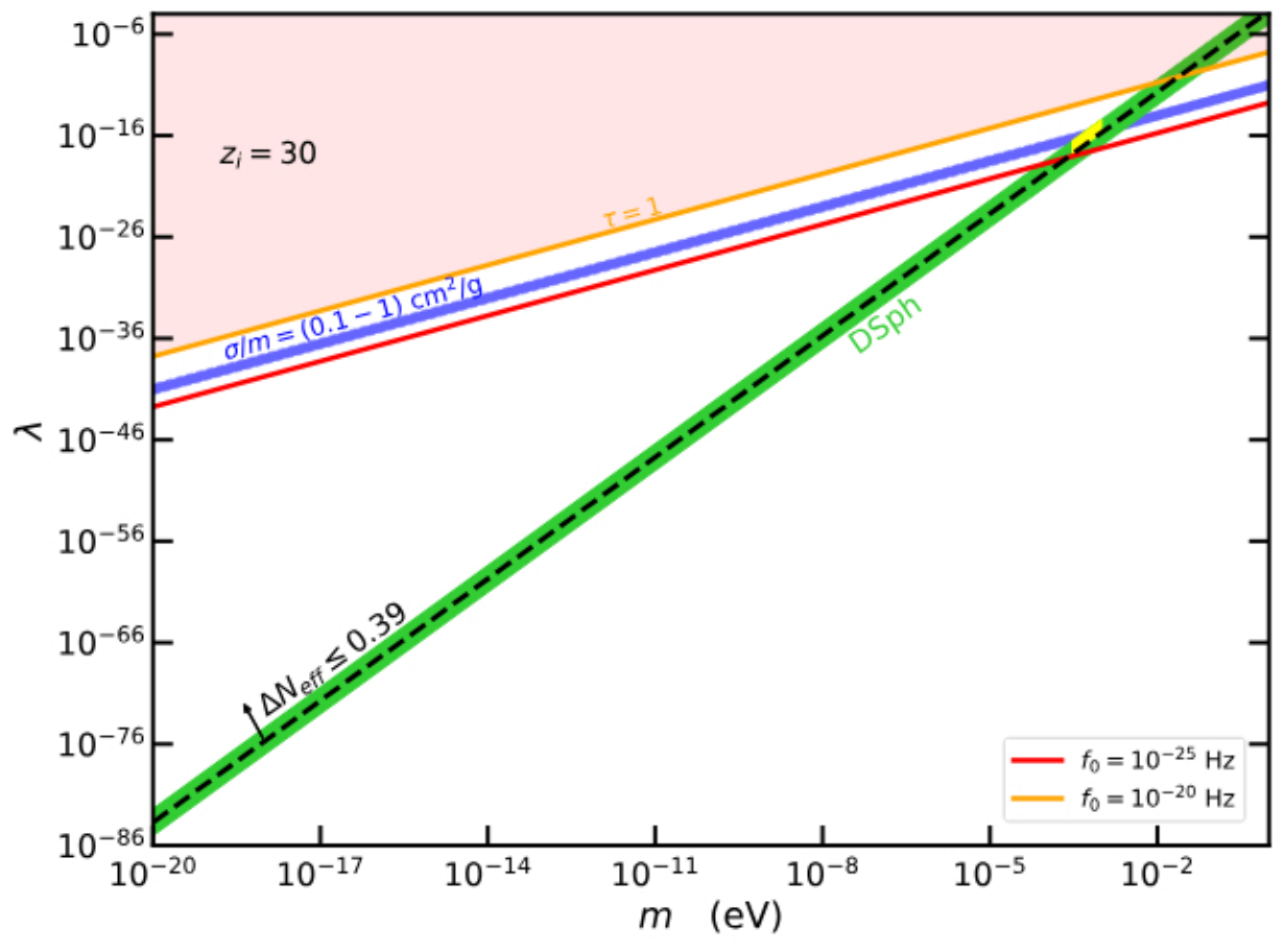}
  \label{fig:scalar halos}
\end{minipage}%
\begin{minipage}{.5\textwidth}
  \centering
  \includegraphics[width=\linewidth]{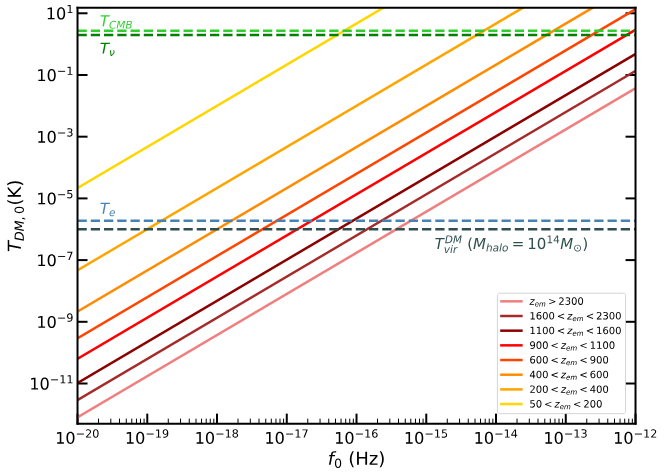}
  \label{fig:test2}
\end{minipage}
\vspace{-0.6 cm}

 \caption{\textit{Left:} Mass and coupling parameter space. Solid lines represent the values by which $\tau = 1$ at observed frequencies $ f_0 =10^{-20}$ Hz (orange)  and $ f_0 =10^{-25}$ Hz (red). Additional existing constraints are plotted. In the red-shaded region $\tau \gg 1$.  In yellow, physically preferred regions are highlighted. The original plot is from ref. \cite{BEC}. \textit{Right:} Parameter space for the temperature of DM and frequency of observation. We take physically preferred values $(m, \lambda) \approx (10^{-4}\;\text{eV}, 10^{-19})$. The plots set upper constraints on $T_{DM}$ at present for a given frequency of observation and differenet redshifts of emission $z_{em}$.}
\label{fig: fig1}
\end{figure}

We present in this section the constraints that could arise in the hypothetical case we detected GWs of a frequency $f_0$. This detection would set constraints in the optical depth, requiring $\tau \le 1$. From \eqref{UL} and \eqref{TauPS}, one finds
\begin{equation}
    \tau_{\text{halos}} = \frac{ 5.5\cdot10^{-50}\lambda^2}{m^3\, f_0^2}\int_{0}^{z_i}\int_{10^5M_\odot}^{5\cdot10^{14}M_\odot}M^2(1+z)^3\
    \frac{dn_h}{dM}\text{d}M\text{d}z= 3.3\cdot10^{-25}\frac{\lambda^2}{m^3\, f_0^2}.    
\end{equation}
 In the left panel of figure \ref{fig: fig1} we have plotted the values of $\lambda$ and $m$ where $\tau = 1$ for $ f_{0} = 10^{-20}$ and $10^{-25}$ Hz. For comparison, We have also plotted other existing constraints, discussed in detail in \cite{BEC}. In yellow, we have highlighted the physically preferred values according to existing bounds, $(m, \lambda) \approx (10^{-4}\;\text{eV}, 10^{-19})$. In the event of detecting a GW of frequency $f_0$ emitted at $z_i$, the red shaded region would be forbidden as $\tau \gg 1$. The arising constraints are less stringent than existing ones for any frequency $f_0 > 10^{-20}$ Hz.\\
\vspace{-0.15 cm}

 For the background absorption, we don't know the temperature of DM. Instead, we can fix the mass $m$ and the coupling $\lambda$ to its physically preferred values and study the $T_{DM,0}- f_0$ parameter space.  For $z_{em} \gg1$,
 \vspace{-0.3 cm}
\begin{equation}
     \tau_{\text{bg}}  \approx 8\cdot10^{-24}\frac{T_{DM,0}^{3/2}\;\lambda^2 \;z_{em}^{13/2}}{m^{9/2}\; f_0^2}.
\end{equation}
\vspace{-0.2 cm}

In the right panel of figure \ref{fig: fig1}, solid lines indicate $\tau_{\text{bg}} = 1$ for each $z_{em}$. Above these, the DM would be too hot for the medium to be transparent to GWs. For comparison, the temperature of IGM electrons, CMB photons and virialized DM are plotted. It is difficult to quantify how stringent these constraints are, as virial temperatures are already very small for particles of such low masses.
\vspace{-0.2 cm}

\vspace{-0.7 cm}

\section{Final remarks}
\vspace{-0.15 cm}
We have considered what opportunities (IB of) GWs offer to probe models of SIDM in the particular case of ULDM. The arising constraints are less stringent than existing ones, even for unrealistically small frequencies of observation. However, we must point out that this was a very simple analysis, where we imposed $\tau <1$ for a single scattering process - i.e considering this to be the \textit{only} source of GWs absorption. A rigorous treatment would require a full study of all existing sources of absorption, and the corresponding contribution of each channel to $\tau$.  To this purpose, we have developed analytical expressions for $\tau$ in both IGM and galaxies valid for any non-relativistic 2-2 scattering. To our knowledge, this is the first time expressions of such kind are presented in the literature. Further and detailed work is necessary to explore the full potential of this approach to GWs physics.

\newpage
\begin{acknowledgments}
This work was a part of my Master's thesis. Thanks D. Blas for the supervision of the project. This article is based upon work from COST Action COSMIC WISPers CA21106, supported by COST (European Cooperation in Science
and Technology)
\end{acknowledgments}

\end{document}